\documentclass[letterpaper,showpacs,preprintnumbers,amsmath,amssymb,nofootinbib,twocolumn,superscriptaddress,notitlepage,prl]{revtex4-1}

\usepackage{graphicx}
\usepackage[usenames,dvipsnames]{color}
\usepackage[
colorlinks=true,       
linkcolor=red,          
citecolor=blue,        
filecolor=magenta,      
urlcolor=cyan  
]{hyperref}

\usepackage{amsmath}
\usepackage{mathtools}
\usepackage{bbm}
\usepackage{amsfonts}
\usepackage{amssymb}
\usepackage{latexsym}
\usepackage{graphicx}
\usepackage[english]{babel}
\usepackage{multirow}
\usepackage{float}
\usepackage{url}
\usepackage{slashed}
\usepackage{xcolor}

\newcommand{\be}{\begin{equation}}
\newcommand{\ee}{\end{equation}}
\newcommand{\ba}{\begin{array}}
\newcommand{\ea}{\end{array}}
\newcommand{\bea}{\begin{eqnarray}}
\newcommand{\eea}{\end{eqnarray}}

\newcommand{\nn}{\nonumber}

\usepackage{ulem,fancyvrb}
\usepackage{xcolor} 
\newcommand{\zb}[1]{{\color{blue} {#1}}}

\begin{document}

\title{A vector leptoquark interpretation 
of the muon $g-2$ and $B$ anomalies}

\author{Mingxuan Du} 
\affiliation{Department of Physics, Nanjing University, Nanjing 210093, China}

\author{Jinhan Liang} 
\affiliation{Department of Physics, Nanjing University, Nanjing 210093, China}

\author{Zuowei Liu} 
\affiliation{Department of Physics, Nanjing University, Nanjing 210093, China} 
\affiliation{CAS Center for Excellence in Particle Physics, Beijing 100049, China} 

\author{Van Que Tran} 
\affiliation{Department of Physics, Nanjing University, Nanjing 210093, China}

\begin{abstract}

We show that a single vector leptoquark can explain 
both the muon $g-2$ anomaly recently measured by 
the Muon g-2 experiment at Fermilab, 
and the various B decay anomalies, 
including the $R_{D^{(*)}}$ and 
$R_{K^{(*)}}$ anomalies which have 
been recently reported 
by the LHCb experiment. 
In order to provide sizeable positive 
new physics 
contributions to the muon $g-2$, 
we assume that the vector leptoquark particle 
couples to both left-handed and right-handed fermions 
with equal strength.
Our model is found to satisfy 
the experimental constraints 
from the large hadron collider.

\end{abstract}

 \maketitle

{\it Introduction:--}Recently, 
the Muon g-2 experiment at Fermilab, 
E989, has announced its 
new measurement on the anomalous 
magnetic moment of muon
\be
a_\mu = {g_\mu-2 \over 2}.
\ee
The result of the 
previous muon g-2 experiment at BNL, 
E821, is $3.7 \sigma$ from the standard 
model (SM) expectation. 
The new E989 Run 1 data give  
a smaller $a_\mu$ with better accuracy
which is $3.4\sigma$ 
away from the SM value. 
The combined results of the Fermilab 
E989 Run 1 data and the BNL E821 data give rise to  
a slightly smaller $a_\mu$ with a better precision, 
yielding a $4.24\sigma$ deviation from the SM value 
\cite{fermilab-muon-g2,Aoyama:2020ynm}
\be
\Delta a_\mu = a_\mu^{\rm Exp} 
- a_\mu^{\rm SM} = (251 \pm 59)\times 10^{-11}. 
\ee

Recently, various $B$ meson decays have shown 
significant deviations from the SM predictions, 
most of which are related to muon final states. 
The lepton flavour universality in $B$ meson decays 
can be tested 
by measuring the ratios of the $b\to s \ell \ell$ transitions 
\bea
R_K = \frac{{\rm BR}(B^+ \to K^+ \mu^+ \mu^-)}{{\rm BR}(B^+ \to K^+ e^+ e^-)}, \\
R_{K^{*}} = \frac{{\rm BR}(B^0 \to K^{*0} \mu^+ \mu^-)}{{\rm BR}(B^0 \to K^{*0}  e^+ e^-)},
\eea
and the ratios of the $b\to c \ell \bar{\nu_{\ell}}$ decays
\be
R_{D^{(*)}} = \frac{{\rm BR}(B \to D^{(*)} \tau \bar{\nu})}
{{\rm BR}(B \to D^{(*)} \ell \bar{\nu})},
\ee
where $\ell = e, \mu$. 
Recently, LHCb has updated the 
measurement on $R_K$ 
\cite{Aaij:2021vac} 
\be
R_K^{\rm {Exp}} = 0.846^{+0.042+0.013}_{-0.039-0.012}, 
\ee  
in the region of $q^2 = [1.1, 6]$ GeV$^2$, 
which is 3.1$\sigma$ away from  
the SM prediction \cite{Bobeth:2007dw, Bordone:2016gaq}  
$R_K^{\rm SM} = 1.0003\pm0.0001$.
The LHCb has also reported the results of $R_{K^*}$ \cite{Aaij:2017vbb}, 
\bea
R_{K^{*}}^{\rm {Exp}} = 0.660^{+0.110}_{-0.070} \pm {0.03}, \\
R_{K^{*}}^{\rm {Exp}} = 0.685^{+0.113}_{-0.069} \pm {0.05},
\eea
in the regions of $q^2 = [0.045, 1.1]$ GeV$^2$ 
and $q^2 = [1.1, 6]$ GeV$^2$ respectively, 
indicating $2.2 \sigma$ and $2.4 \sigma$ deviations from 
the SM predictions, which are
$R_{K^{*}}^{\rm SM} = 0.92\pm 0.02$ and
$R_{K^{*}}^{\rm SM} = 1.00 \pm 0.01$ in these two regions.
Together with the recent results from Belle \cite{Abdesselam:2019dgh}, 
the world averages for $R_{D^{(*)}}$ measurements
are \cite{Amhis:2019ckw}
$R_D^{\rm {Exp}} = 0.340\pm 0.030$ and 
$R_{D^*}^{\rm {Exp}} = {0.295} \pm 0.014$,
whereas the SM expectations are 
\cite{Amhis:2019ckw}
$R_D^{\rm SM} = {0.299\pm 0.003 } $ 
and $R_{D^*}^{\rm SM} = {0.258 \pm 0.005}$, 
{yielding a $\sim 3 \sigma$ deviation 
when these two are combined.

Recently, LHCb has also released 
its measurement on 
$B_s \to \mu^+\mu^-$ with 
the full run 2 data \cite{LHCb:Bsmumu}
\be
\rm{BR}(B_s \to \mu^+\mu^-) = (3.09^{+0.46+0.15}_{-0.43-0.11}) \times 10^{-9}. 
\ee
By updating the world average of $B_s \to \mu^+\mu^-$ branching ratio and the correlated $B^0 \to \mu^+\mu^-$, 
Ref.~\cite{Altmannshofer:2021qrr} found a $2.3 \sigma$ deviation 
from the SM prediction.

In this paper, we use a single 
vector leptoquark (LQ) 
to explain both the muon $g-2$ anomaly 
and the various B decay anomalies, 
in particular the 
$R_{K^{(*)}}$ and $R_{D^{(*)}}$  
anomalies, 
which  have been measured more precisely 
recently and shown significant deviations 
from the SM. 
LQs are new physics particles 
that couple simultaneously to a lepton 
and a quark; see e.g.\ \cite{Dorsner:2016wpm} 
for a recent review. 
LQ models have been proposed to 
explain various $B$ anomalies. 
A single scalar LQ 
has been proposed 
to explain both the muon $g-2$ 
and the $B$ anomalies \cite{Bauer:2015knc} 
\footnote{Models with multiple scalar leptoquarks  
have also been explored  
\cite{Chen:2017hir,Nomura:2021oeu}.}.
However, recently, 
Ref.\ \cite{Angelescu:2021lln} 
shows that it is difficult for 
a single scalar LQ to explain both $R_{K^{(*)}}$ 
and $R_{D^{(*)}}$ anomalies
because the measured $R_{K^{(*)}}$ values are smaller than the 
SM expectations whereas $R_{D^{(*)}}$ are larger.
It is also found that
a vector LQ 
that transforms as $({\bf 3},{\bf 1},2/3)$ 
under the standard model gauge group 
$SU(3)_c \times SU(2)_L \times U(1)_Y$
can explain both 
$R_{K^{(*)}}$ 
and $R_{D^{(*)}}$ anomalies \cite{Angelescu:2021lln}. 
In this study, we 
further show that such a vector LQ model can also 
explain the muon g-2 anomaly, in addition 
to the B meson decay anomalies which 
satisfying various LHC constraints. 
Our model is different from that 
considered in Ref.\ \cite{Angelescu:2021lln} 
because we consider the couplings to both 
left-handed and right-handed fermions, 
whereas only left-handed couplings are 
assumed in Ref.\ \cite{Angelescu:2021lln}. 
Our analysis shows that both left-handed 
and right-handed couplings between the 
$U_1$ boson and the SM fermions are essential 
for a sizeable new physics contribution 
to explain the muon g-2 anomaly.

{\it Leptoquark model:--}We consider the 
vector LQ 
$U_1=({\bf 3},{\bf 1},2/3)$ 
that couples to both left-handed and 
right-handed standard model fermions. 
The interaction Lagrangian between the vector LQ $U_1^\mu$ 
and SM fermions in the weak basis is given by 
\be
\label{eq:Lag}
{\cal L} = x_{ij}^L \bar{Q}_L^i \gamma_\mu L_L^{j} U_1^\mu 
+ x^R_{ij} \bar{d}_R^i\gamma_\mu \ell_R^j U_1^\mu + h.c.,  
\ee
where  
${Q}_L^{\rm T} = (u_L,d_L)$ is the left-handed quark doublet, 
${L}_L^{\rm T} = (\nu_L,\ell_L)$ is the 
left-handed lepton doublet, 
${d}_R$ ($\ell_R$) is the right-handed 
quark (lepton), 
$x^L_{ij}$ and $x^R_{ij}$ are couplings 
with $i$ and $j$ as the generation index of the SM fermions.
Here we assume that the down quarks and the 
charged leptons are diagonal. 
Rotating the up quark fields to the mass basis, 
the interaction Lagrangian becomes 
\be
{\cal L} = (V x^L)_{ij} \bar{u}^i_L \gamma_\mu \nu_L^{j} U^\mu_1 
+ x^{L}_{ij} \bar{d}_L^i \gamma_\mu \ell_L^{j} U_1^{\mu}
+ x^R_{ij} \bar{d}_R^i\gamma_\mu \ell_R^j U_1^{\mu}
+ h.c.\zb{,}
\label{eq:L:CKM}
\ee
where V is the CKM matrix. 
We do not consider couplings with right-handed 
neutrinos. 
We assume that the coupling between 
the vector LQ $U_1$ and the SM gauge bosons  
(photon and gluon) are standard gauge 
couplings. 
The $U_1$ coupling to photon contributes to  
the muon $g-2$. 
The $U_1$ coupling to gluons are 
important for the production cross section 
of $U_1$ at the LHC.

\begin{figure}[htbp]
\begin{center}
\includegraphics[height=3cm,width=0.2\textwidth]{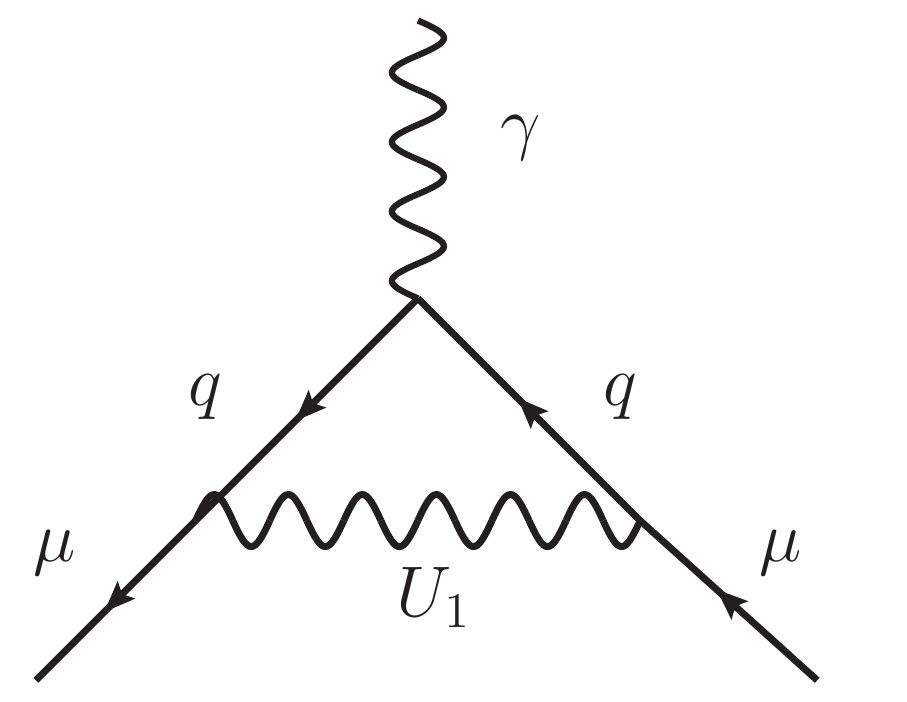}
\includegraphics[height=3cm,width=0.2\textwidth]{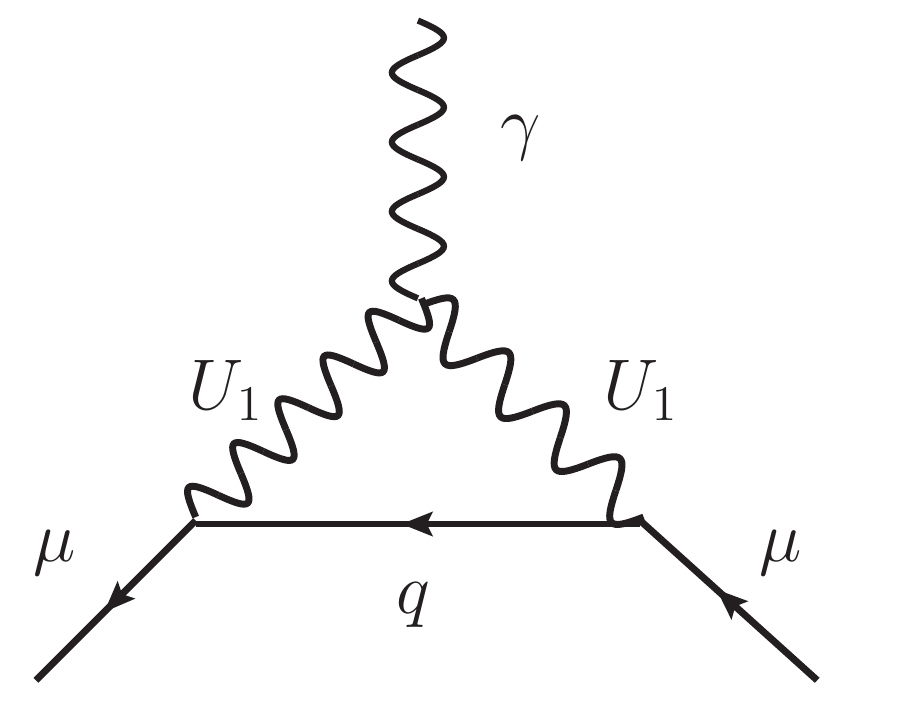}
\caption{New contributions to muon $g-2$ from 
the vector LQ $U_1$. 
} 
\label{fig:feyn-muon}
\end{center}
\end{figure}

{\it Muon g-2 anomaly}:--
In our vector LQ model,
the main new physics contributions 
to the muon $g-2$ 
arise from the loop 
diagrams shown in Fig.\ \ref{fig:feyn-muon}. 
The vector LQ contributions 
are given by
\cite{Leveille:1977rc, Jegerlehner:2009ry, Lindner:2016bgg, Biggio:2016wyy}
\bea
\Delta a_{\mu} = {-}\frac{{N_c Q_q}}{8\pi^2}
\int_0^1 dx \frac{f^2_{v} F_{v}(x,\epsilon)+ f^2_{a} F_{a}(x,\epsilon)}
{(1-x)(\lambda^{-2}- x) +\epsilon^2  x} \nonumber\\
{+}\frac{{N_c Q_{U_1}}}{8\pi^2}
\int_0^1 dx \frac{f^2_{v} F_{v}^{\prime}(x,\epsilon)+ f^2_{a} F_{a}^{\prime}(x,\epsilon) }{(1-x)(\epsilon^2 -x)+ \lambda^{-2}x},
\label{eq:muon-g-2}
\eea
where $N_c=3$ is color number, $Q_q$ is the 
electric charge of the SM quark, 
and $Q_{U_1}=2/3$  
is electric charge of $U_1$, 
$\epsilon=m_{q}/m_{\mu}$, 
$\lambda = m_{\mu}/m_{U_1}$, 
$f_v =  (x^R+x^L)/2$, 
$f_a = (x^R-x^L)/2$, 
$F_{v}(x,\epsilon)  =  2x(1-x)\left(x-2(1-\epsilon)\right)
+\lambda^2(1-\epsilon)^2 x^2(1+\epsilon-x)$, 
$F_{v}^{\prime}(x,\epsilon) = -2x^2(1+x-2\epsilon) + \lambda^2(1-\epsilon)^2 x(1-x)(x+\epsilon)$, 
$F_{a}(x,\epsilon)=F_{v}(x,-\epsilon)$, and 
$F_{a}^{\prime}(x,\epsilon)=F_{v}^{\prime}(x,-\epsilon)$.

We consider four different types of 
couplings: 
$x^L = x^R$, 
$x^L = - x^R$, 
$x^L = 0$, and  
$x^R = 0$; 
we find that only the 
$x^L = x^R$ case can give rise to 
a sizeable positive new physics 
contribution to the 
muon $g-2$. 
Thus we assume $x^L = x^R$ 
hereafter. 
We denote $x\equiv x^L = x^R$. 
We note that the $U_1$ boson 
has pure vector couplings to fermions 
in the $x^L = x^R$ case. 
To explain the muon $g-2$, 
$R_{K^{(*)}}$ {and $R_{D^{(*)}}$} anomalies, 
we assume nonzero 
$x_{s\mu}$,  
$x_{b\mu}$ 
{and $x_{b\tau}$} values.

\begin{figure}[htbp]
\begin{centering}
\includegraphics[width=0.4\textwidth]{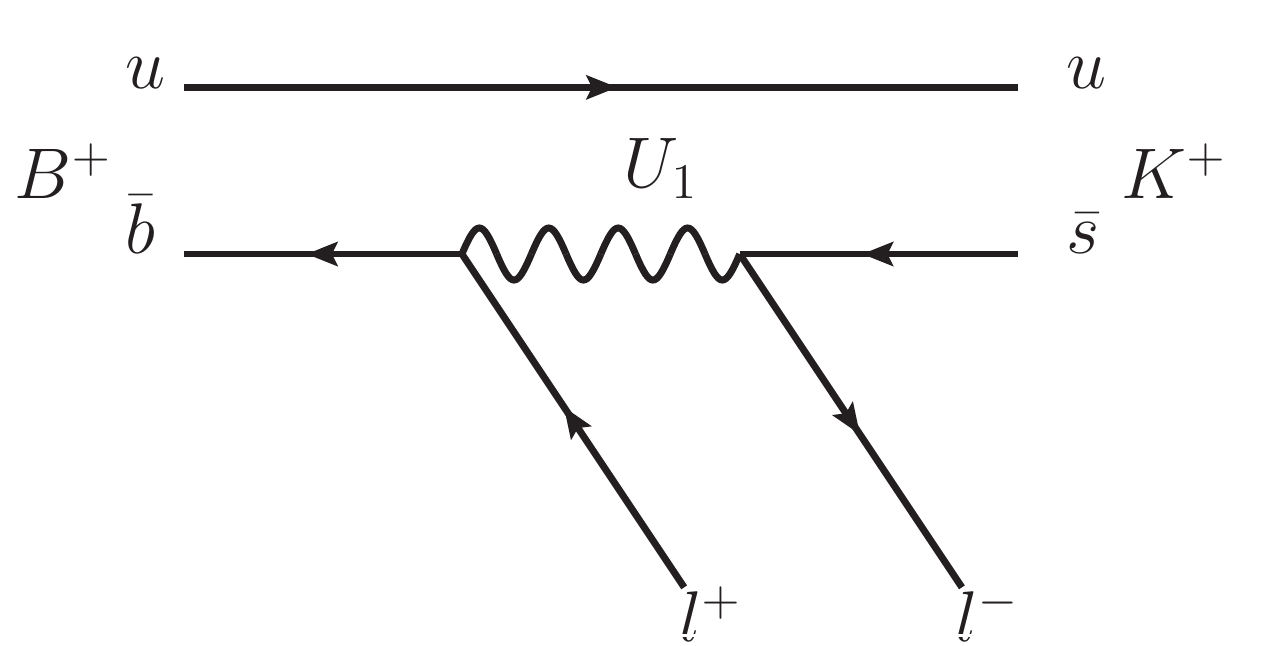} 
\caption{New contributions to  
the $B^+ \to K^+ l^+ l^-$ process due to 
the vector LQ $U_1$.
}
\label{fig:feyn-b2k}
\end{centering}
\end{figure}

{\it $R_{K^{(*)}}$ anomaly:--}The  
semileptonic process $b \to s \ell^+\ell^-$ 
is responsible for the 
$B^+ \to K^+ \ell^+ \ell^-$ decay. 
The new physics contributions due to the 
new vector LQ $U_1$ are  
shown in Fig.~\ref{fig:feyn-b2k}. 
The effective Hamiltonian describing the 
$b \to s \ell^+\ell^-$ process 
can be parameterized as follows 
\cite{Grinstein:1987vj,Buchalla:1995vs,Buras:1998raa,Cornella:2019hct}
\be
{\cal H} = - \frac{G_F \alpha}{\sqrt{2}\pi}
V_{tb}V_{ts}^{*} \sum_i{{\cal C}_i {\cal O}_i + h.c.\zb{,}}
\ee
where ${\cal O}_i$ 
are the four-fermion operators,  
${\cal C}_i$ are the 
Wilson coefficients, 
$G_F$ is the Fermi constant, 
$\alpha \sim 1/137$,  
$V_{tb}$ and $V_{ts}$ are the CKM matrix element. 
The operators that are of importance 
to the $R_{K^{(*)}}$ analysis in our model 
include 
${\cal O}_{9, 9'}^{\ell\ell} = 
\bar{s}\gamma_\mu P_{L,R} b (\bar{\ell}\gamma^{\mu}\ell)$, and  
${\cal O}_{10, 10'}^{\ell\ell} = \bar{s}\gamma_\mu P_{L,R} b (\bar{\ell}\gamma^{\mu}\gamma_5\ell)$, 
where $\ell = e, \mu, \tau$ and 
$P_{R,L} = (1\pm \gamma_5)/2$.\footnote{Operators 
such as ${\cal O}_{S, S',P,P'}$ also arise 
in the $x^L = x^R$ case. 
These operators are neglected 
in our analysis, because they 
do not give rise to significant contributions to $R_{K^{(*)}}$ as the 
${\cal O}_{9, 9',10,10'}$ operators, 
as shown in Ref.~\cite{Ghosh:2017ber}.}
The Wilson coefficients can be parameterized as 
${\cal C}_i = {\cal C}_i^{\rm SM} + \Delta {\cal C}_i$ 
where ${\cal C}_i^{\rm SM}$ represents 
the SM contribution 
and $\Delta{\cal C}_i$ denotes the new physics contributions. 
For the vector LQ
$U_1$ model where nonzero 
$x_{b\mu}$ and $x_{s\mu}$ values 
are introduced, 
the new physics 
contributions to the Wilson coefficients 
are given by 
\be
\Delta {\cal C} \equiv \Delta {\cal C}_9^{\mu\mu} = \Delta {\cal C}_{9'}^{\mu\mu}= \Delta {\cal C}_{10'}^{\mu\mu} = - \Delta {\cal C}_{10}^{\mu\mu}, 
\label{eq:xL:RK_new}
\ee
where 
\be
\Delta {\cal C} = -\frac{\pi v^2} {V_{tb}V_{ts}^*\alpha} \frac{x_{s\mu} x_{b\mu}^*}{m^2_{U_1}}.
\label{eq:xL:RK}
\ee
where $v=246$ GeV.  

For the $R_K$ calculations, operators ${\cal O}_{i}$ 
and ${\cal O}_{i'}$ yield nearly 
the same effects, for $i = 9,10$.
Because our model predicts 
$\Delta {\cal C}_{9}^{\mu\mu} =  \Delta {\cal C}_{9'}^{\mu\mu}$ 
and 
$\Delta {\cal C}_{10}^{\mu\mu} = - \Delta {\cal C}_{10'}^{\mu\mu}$ 
 in the $x^L = x^R$ case, 
the effects on $R_K$ due to  
${\cal O}_{10}$ cancel with 
${\cal O}_{10'}$, 
and we only need to consider the effects due to the
operator ${\cal O}_{9}$ 
with the coefficient $2\Delta {\cal C}$. 
We determine the $\Delta {\cal C}$
value needed to interpret the recent 
LHCb $R_K$ result 
\cite{Aaij:2021vac} 
using the 
theoretical 
analysis on $R_K$ in  
Ref.~\cite{Ghosh:2017ber} 
and find that 
\be
{\Delta {\cal C} (R_K) \simeq -0.35 \pm 0.11}
\label{eq:RK-fit}
\ee

For the $R_{K^{*}}$ calculations, because operators ${\cal O}_{i}$ 
and ${\cal O}_{i'}$ 
give rise to nearly the same 
contributions for $i = 9,10$ but with 
opposite signs, 
the net effects in our model 
can be approximated by the 
operator ${\cal O}_{10}$ 
with the coefficient -$2\Delta {\cal C}$. 
The required $\Delta {\cal C}$ value 
for the $R_{K^{*}}$ \cite{Aaij:2017vbb}
interpretation is also obtained 
via the analysis in Ref.~\cite{Ghosh:2017ber}
\be
{\Delta {\cal C} (R_{K^{*}}) \simeq -0.63^{+0.28}_{-0.22}}
\label{eq:RKs-fit}
\ee
Thus one can explain both 
$R_{K}$ and $R_{K^{*}}$ anomalies 
within the 1 $\sigma$ (2 $\sigma$) error corridor   
in the range $-0.46 < \Delta {\cal C} < -0.35$ 
($-0.57 < \Delta {\cal C} < -0.13$).\footnote{Our results are similar 
to the global fit (for the $x^R = 0$ case)
$\Delta {\cal C} (R_K)  = -0.41 \pm 0.09$ 
where both $R_K$ and $B_s \to \mu^+ \mu^-$ 
measurements are taken into account 
\cite{Angelescu:2021lln}.}

{\it $R_{D^{(*)}}$ anomaly:--}
Here we compute $R_{D^{(*)}}$ 
for the LQ model we consider. 
The relevant low-energy EFT Lagrangian 
describing
the $b\to c\ell \nu$ 
is given by \cite{Shi:2019gxi,Murgui:2019czp,Alonso:2015sja}
\be
\label{eq:lag:R_D}
{\cal L} = - 2 \sqrt{2} G_{F}V_{c b}\Big[\left(1+y^{\ell}_{V_{L}}\right) \mathcal{O}_{V_{L}}  
+y^{\ell}_{S_{R}} \mathcal{O}_{S_{R}} \Big] + h.c.\zb{,}
\ee
where $y^{\ell}_{V_{L}}$ and $y^{\ell}_{S_{R}}$ 
are the Wilson coefficients
and the four-fermion operators are given by 
\be
\mathcal{O}_{V_{L}}=\left(\bar{c}_L \gamma^{\mu} b_{L}\right)\left(\bar{\ell}_{L} \gamma_{\mu} \nu_{ L}\right), 
\quad \mathcal{O}_{S_{ R}}=\left(\bar{c}_L b_{ R}\right)\left(\bar{\ell}_{R} \nu_{ L}\right).
\ee
The new physics contributions 
to the Wilson coefficients at tree level from the
vector LQ model can be obtained as 
\bea
\label{eq:y-VL}
y^{\ell}_{V_L} &=& -\frac{1}{2} y^{\ell}_{S_R} = \frac{v^2}{2 V_{cb}} \frac{(V x)_{c\ell'} (x_{b\ell})^*}{m^2_{U_1}} \nn \\
&=& \frac{v^2}{2 V_{cb} m_{U_1}^2} (V_{cb} x_{b\ell'} + V_{cs} x_{s\ell'}) (x_{b\ell})^*, 
\eea
where the sum over $\ell'$ for generations of the neutrino is implicit 
on the right hand side. 
We assume that the non-zero matrix elements 
in our model are 
$x_{s\mu}$,  
$x_{b\mu}$ 
{and $x_{b\tau}$} values. 
In our numerical analysis, we set $x_{b\tau} = 5 x_{b\mu}$
so that the new physics contributions to the 
$b\to c\mu \nu$ transitions are much smaller than the 
new physics contributions to the $b\to c\tau \nu$ 
transitions. Thus  
We neglect the new physics contributions to 
the $b\to c\mu \nu$ transitions 
and adopt the analysis in Ref.~\cite{Shi:2019gxi} 
to obtain 
\be
\label{eq:yVL-bt}
y^{\tau}_{V_L} = 0.08^{+0.03}_{-0.03},\quad   y^{\tau}_{S_R} = -0.05^{+0.09}_{-0.10}, 
\ee
which further give rise to 
\be
\label{eq:bf:yL}
y^{\tau}_{V_L} = -\frac{1}{2} y^{\tau}_{S_R} = 0.065 \pm 0.027, 
\ee
when the two fits are combined. 
In order to explain the relation 
$R_{K^{(*)}}^{\rm Exp} < R_{K^{(*)}}^{\rm SM}$, 
the matrix elements $x_{s\mu}$ and $x_{b\mu}^{*}$  
have to be opposite in sign. 
Taking all matrix elements to be real, 
and setting both $x_{b\mu}$ and $x_{b\tau}$ positive 
and $x_{s\mu}$ negative, 
in order to explain  
$R_{D^{(*)}}^{\rm Exp} > R_{D^{(*)}}^{\rm SM}$, 
the condition $|x_{s\mu}| < (x_{b\mu} + x_{b\tau}) V_{cb}/V_{cs}$ 
has to be satisfied. 
Thus in our vector LQ model, both $R_{K^{(*)}}$ 
and $R_{D^{(*)}}$ anomalies can be explained 
simultaneously.

{\it LHC constraints:---}The LQs can be either singly produced 
or pair-produced 
at hadron colliders \cite{Diaz:2017lit, Dorsner:2018ynv}. The singly produced LQ is accompanied by 
a lepton in the final state, 
$qg \to q \to U_1 \ell$; 
the pair-production process can occur  
via either the gluon–gluon fusion process 
or the quark–antiquark annihilation process   
(mediated by either a t-channel lepton
or an s-channel photon/gluon).

Searches for these direct productions of leptoquarks 
have been carried out 
at the LHC for various final states
\cite{Aaboud:2019bye, Aad:2021rrh, Aad:2020iuy,Aad:2020jmj, CMS:2018}. 
From these searches, one can derive 
the collider exclusion bounds 
on a given leptoquark mass as a function 
of its branching ratio (denoted by $\beta$) into 
a specific fermion final state.

Using the recent LHC data, 
a recasting result from Ref.~\cite{Angelescu:2021lln}
shows the current limits for 
the vector LQs. 
In particular, for pair-produced LQs decaying into the final states of 
$b\bar{b} \tau \bar \tau$, $t\bar{t} \tau \bar \tau$, 
$j j \mu\bar\mu$, $b\bar{b} \mu \bar \mu$, 
$t\bar{t} \mu \bar \mu$, $j j \nu \bar \nu$, 
$b\bar{b} \nu \bar \nu$ and $t\bar{t} \nu \bar \nu$, 
the lower limits on its mass are 
$1.5$ TeV, $2.0$ TeV,  $2.3$ TeV, $2.3$ TeV, 
$2.0$ TeV, $1.8$ TeV, $1.8$ TeV and $1.8$ TeV, respectively, 
assuming $\beta = 1$ \cite{Angelescu:2021lln}.
The limits from
$b\bar{b} \tau \bar \tau$,
$j j \mu\bar\mu$, $b\bar{b} \mu \bar \mu$, 
$j j \nu \bar \nu$, and $t\bar{t} \nu \bar \nu$ final states 
can be applied for the $U_1$ model. 
The most stringent constraints are 
from $j j \mu\bar\mu$ and $b\bar{b} \mu \bar \mu$ final states.
For the pair of LQs decaying into 
different quark-lepton final states case,  
i.e., $U_1 \to b \bar{\tau}, t\bar{\nu}$, 
the lower limit on the vector LQ is about
$1.7$ TeV \cite{Sirunyan:2020zbk}.

However, these limits are under the assumption that 
the branching fraction $\beta = 1$ and 
the interaction between vector leptoquark
and gluons is described by $\kappa g_s U_{1\mu}^{\dagger} G^{\mu\nu} U_{1\nu}$ 
with $g_s$ is the strong coupling and $\kappa = 1$. 
Thus if one tunes $\beta$ and $\kappa$
to be small, the limits from LHC direct production searches can be 
significantly weakened.
Indeed, once switching off the 
the interaction between the 
vector LQ and gluons, i.e., $\kappa = 0$, 
the quark–antiquark annihilation process becomes dominant; 
because the quark–antiquark annihilation for the
vector LQ pair production is typically 
smaller than the gluon fusion, 
the limit on the vector LQ mass is reduced to 
$\sim 1.3$ TeV 
for the $U_1 \to b \bar{\tau}, t\bar{\nu}$ searches
\cite{Sirunyan:2020zbk}.

The searches for the singly produced LQs 
have also been performed at CMS 
using data with ${\cal L} = 137~ {\rm{fb}}^{-1}$ \cite{Sirunyan:2020zbk}. 
The limits on the LQ mass as a function of the LQ-quark-lepton coupling $x$ 
have been derived; for LQs coupled to 
the third-generation fermions with coupling $x = 1.5$, 
the LQ mass $m_{U_1} \lesssim 1.2$ TeV is excluded.

LQs can also be searched for 
in the $q q \to \ell \ell$ process 
at the LHC 
in which the LQ particle serves as a 
t-channel mediator.
Thus LQs are constrained by the high $p_T$ 
resonance searches via $pp \to \ell \ell^{(\prime)}$ 
and $p p \to \ell \nu$ recently performed at the LHC 
\cite{Faroughy:2016osc, Schmaltz:2018nls, Greljo:2017vvb, Afik:2019htr, Angelescu:2020uug, Angelescu:2021lln, Greljo:2018tzh, Marzocca:2020ueu}. 
A recent analysis from Ref.~\cite{Angelescu:2021lln}
used data from \cite{Aad:2020zxo} and \cite{CMS:2019tbu} 
to set upper bounds on the
couplings $x_{b\mu} \lesssim 0.7$, $x_{s\mu} \lesssim 0.5$, $x_{b\tau} \lesssim 1.0$ and $x_{s\tau} \lesssim 0.7$ for the vector LQ mass below $1$ TeV in $U_1$ model.
The limit on $x_{b\mu}$ as a function of $m_{U_1}$ is shown in Fig.~\ref{fig:mU1-xbmu}.

\begin{figure}[htbp]
\begin{centering}
\includegraphics[width=0.4\textwidth]{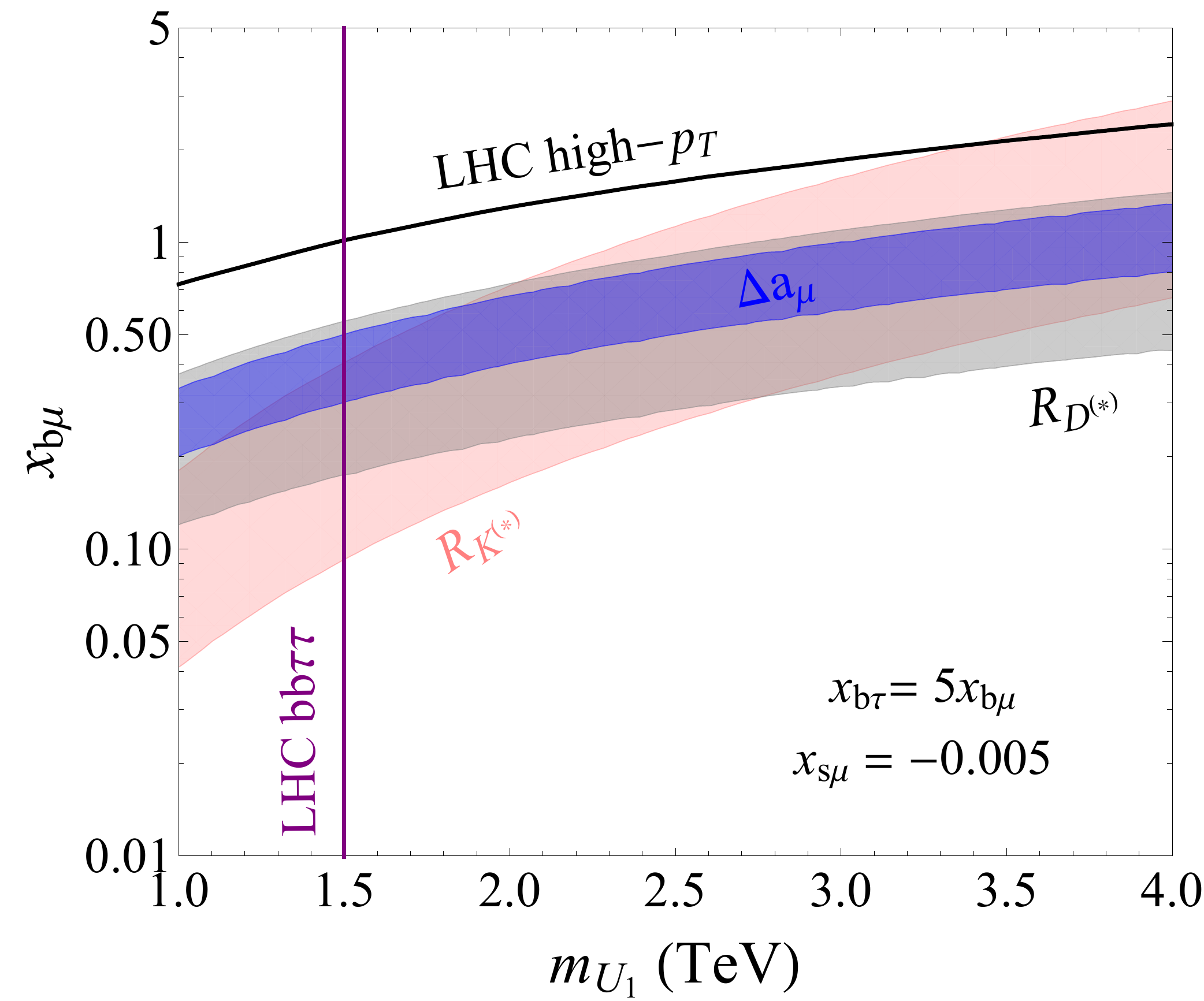} 
\caption{The $2\sigma$ favored regions for
the muon $g-2$ (blue band), $R_{K^{(*)}}$ (pink band) 
and $R_{D^{(*)}}$ (gray band) {on the} 
$(m_{U_1}, x_{b\mu})$ plane. 
Here we set $x_{s\mu} = - 0.005$ 
and $x_{b\tau} = 5 x_{b\mu}$. 
The black solid line indicates the LHC high $p_T$ searches, 
and the vertical purple solid line represents  
the leptoquark searches via pair-production 
at the LHC \cite{Angelescu:2021lln}. 
}
\label{fig:mU1-xbmu}
\end{centering}
\end{figure}

{\it Numerical results:---}Here we present our numerical results. 
We choose  $x_{b_\tau} = 5 x_{b\mu}$ so that 
the new physics contributions to 
the $b \to c\tau \nu$ process 
are larger than those to the $b \to c\mu \nu$ 
process,
leading to $R_{D^{(*)}} > R_{D^{(*)}}^{\rm SM}$. 
Note that varying the $x_{b_\tau}$ coupling 
does not affect the
$R_{K^{(*)}}$ and $\Delta a_{\mu}$ computations. 
We choose $x_{s\mu} = - 0.005$ 
which has an opposite sign to $x_{b\mu}$ 
in order to satisfy the $R_{K^{(*)}}$ anomaly 
which is required to obtain $R_{K^{(*)}}^{\rm Exp} < R_{K^{(*)}}^{\rm SM}$. 
Such a small value of $x_{s\mu}$ also satisfies the condition 
$|x_{s\mu}| < (x_{b\mu} + x_{b\tau}) V_{cb}/V_{cs}$ 
needed for 
a positive contribution to the $R_{D^{(*)}}$.

Fig.\ \ref{fig:mU1-xbmu} shows the 2$\sigma$ 
regions of the parameter space of our vector LQ model, 
for the muon $g-2$, $R_{K^{(*)}}$, and $R_{D^{(*)}}$ anomalies. 
It is remarkable that there exists a large 
parameter space ranging from 1 TeV to above 4 TeV, 
in which one can simultaneously 
explain the muon $g-2$, $R_{K^{(*)}}$ and $R_{D^{(*)}}$ anomalies. 
The LHC constraints including the high $p_T$ 
di-lepton searches and the  pair-produced 
LQ searches are also displayed on 
 Fig.\ \ref{fig:mU1-xbmu}; 
both LHC constraints are adopted from 
Ref.~\cite{Angelescu:2021lln}. 
Because we have $x_{b\tau} = 5 x_{b\mu}$, 
the branching fraction of ($U_{1} \to b \tau$) 
dominates, namely BR$(U_{1} \to b \tau) \sim 1$.   
Thus we adopt the limits in  
the $b\bar b \tau \bar \tau$ final state searches 
for pair-produced LQs  
at the LHC \cite{Aaboud:2019bye, Angelescu:2021lln}, 
which rule out the $U_1$ boson with a mass 
below $1.5$ TeV. 
The LHC high $p_T$ data \cite{Aaboud:2019bye} 
exclude the $x_{b\mu} > 1$ region 
which have already put constraints on 
the 2 $\sigma$ region for $R_{K^{(*)}}$ interpretation 
when $m_{U_1} > 3.5$ TeV. 
We note that the large parameter space in our model, 
from 1.5 TeV to 4 TeV, in which both the 
muon g-2 anomaly and the various B anomalies 
can be explained are allowed by the current 
LHC searches on LQs.

{\it Conclusions:--}We have shown that 
the vector LQ $U_1= ({\bf 3},{\bf 1},2/3)$ can simultaneously 
explain the muon $g-2$ anomaly and 
the $B$ decay anomalies, including the 
$R_{K^{(*)}}$ and $R_{D^{(*)}}$ 
anomalies, while satisfying 
the experimental constraints from 
the large hadron collider,
including searches on singly produced and pair-produced LQs, 
and the high $p_T$ dilepton searches.

We found that in order to provide a 
sizeable positive new physics contribution to the muon g-2 
anomaly while satisfying various LHC constraints, 
the vector LQ $U_1$ has to couple to left-handed 
and right-handed fermions equally, namely 
$x_L = x_R$, so that only  
vector couplings are present in our model. 
Unlike the scalar LQ model, 
the new physics contributions from the vector LQ to 
$R_{K^{(*)}}$ and $R_{D^{(*)}}$ 
can be somewhat adjusted independently 
in the parameter space of our model.  
Thus conditions $R_{K^{(*)}}^{\rm Exp} < R_{K^{(*)}}^{\rm SM}$ and 
$R_{D^{(*)}}^{\rm Exp} > R_{D^{(*)}}^{\rm SM}$ 
can be satisfied with ease. 
The LHC searches for LQ decays set  
stringent constraints on the LQ mass, $m_{U_1} \gtrsim {\cal O}(1)$ TeV. 
The couplings of LQs to quark and lepton 
are strongly constrained to be $x \lesssim {\cal O} (1)$
in the LHC high-$p_T$ searches for LQ mass extending to 
several TeV. 
Taking into account all the LHC constraints, there still exists 
a large parameter space in which the muon $g-2$ anomaly recently reported 
by the Fermilab g-2 experiment and 
the various $B$ decay anomalies including $R_{K^{(*)}}$ and $R_{D^{(*)}}$ 
can be simultaneously explained in the vector leptoquark model 
in the mass range from 1.5 TeV to several TeV.

{\it Acknowledgement:--}
The work is supported in part 
by the National Natural Science 
Foundation of China under Grant No.\ 11775109. 
Van Que Tran would like to thank the Institute of Physics, 
Academia Sinica, 
Taiwan for its hospitality during this work.

\end{document}